# Analytical theory of sub-fs pulse formation in a seeded hydrogen-like plasma-based X-ray laser dressed by an infrared field

I. R. Khairulin[1,2,*], V. A. Antonov[2,3], M. Yu. Ryabikin[2,3], and Olga Kocharovskaya[4]

[1]*N.I. Lobachevsky State University of Nizhny Novgorod,*
*23 Gagarin Avenue, Nizhny Novgorod, 603950, Russia*
[2]*Institute of Applied Physics of the Russian Academy of Sciences,*
*46 Ulyanov Street, Nizhny Novgorod, 603950, Russia*
[3]*Prokhorov General Physics Institute of the Russian Academy of Sciences,*
*38 Vavilov Street, Moscow, 119991, Russia*
[4] *Department of Physics and Astronomy and*
*Institute for Quantum Studies and Engineering,*
*Texas A&M University, College Station, TX 77843-4242, USA*

We derive the analytical theory describing the process of sub-femtosecond pulse formation from a quasi-monochromatic seeding extreme ultraviolet (XUV) radiation, which propagates in active medium of a hydrogen-like plasma-based X-ray laser dressed by a strong infrared laser field. We discuss the ultimate capabilities and limitations of this process on the basis of the derived analytical solution and extensive numerical studies for the case of $Li^{2+}$ plasma-based X-ray laser with a carrier wavelength 13.5nm. We analyze the role of plasma dispersion and find the optimal conditions for the formation of attosecond pulses with the highest contrast. Under the optimal conditions, the influence of amplified spontaneous emission from the active medium is negligible. The peak intensity of the produced XUV pulses can exceed $10^{10}$-$10^{11}$ W/cm$^2$, while the duration of pulses varies in the range of 400-600 as.

## I. INTRODUCTION

Intense attosecond pulses of coherent extreme ultraviolet (XUV) / X-ray field provide a unique combination of high spatial and temporal resolution desirable for the applications in dynamical, element-specific imaging in biochemistry and material science (see reviews on plasma based X-ray lasers [1-4] and attosecond physics [5-9]).

There are three conventional types of sources of ultrashort coherent XUV / X-ray pulses: high harmonic generation (HHG) sources, X-ray free electron lasers (XFELs), and plasma-based X-ray lasers. HHG allows producing pulses with down to tens of attosecond duration in a table-top setup [10-12]. However, in the X-ray range the pulse energy typically remains on a pJ level. XFELs can produce several mJ pulses typically with few-fs duration (recently, the generation of 200 as pulses has been achieved [13]), although with low reproducibility [14-16]. Typically, the pulses suffer from the lack of temporal coherence coming from the shot noise, which is an inevitable part of the FEL amplification process. Furthermore, XFELs are large scale facilities and there are only a few of them in the world. Plasma based X-ray lasers can produce the radiation pulses in the XUV and X-ray ranges with energies up to several mJ, but typically of a rather long picosecond duration [1-4, 17-21] which limits their applications for ultrafast imaging.

Recently, a method has been proposed for the transformation of a picosecond seeding XUV or X-ray radiation into a train of attosecond pulses via the modulation of an active medium of a hydrogen-like plasma-based X-ray laser by a strong optical laser field [22]. The essence of the proposed technique is as follows. Under the action of the modulating field, the upper lasing energy level of the hydrogen-like ions is split into sublevels which oscillate in space and time along with oscillations of the modulating field strength due to the linear Stark effect. Such sub-optical-cycle splitting of the resonant energy level results in a multifrequency response of the ions to the quasi-monochromatic seeding XUV / X-ray field. Due to this, along with amplifica-tion, the spectrum of the resonant field is enriched by a set of sidebands, which are separated from the resonance by even multiples of the frequency of the modulating field. Under the optimal conditions, the sidebands are strong enough and phase-matched, so that the output XUV/ X-ray field becomes a train of nearly bandwidth-limited pulses of either few-fs or attosecond duration (depending on the parameters of the medium and the modulating field). Furthermore, if the plasma is dense for the modulating optical field and the generation of sidebands is suppressed, an active medium of an X-ray laser might be used for the amplification of attosecond pulse trains produced via the HHG process preserving the duration and shape of attosecond pulses [23].

Nevertheless, we have previously shown only the possibility in principle for the attosecond pulse formation from a quasi-monochromatic seeding XUV / X-ray field [22] based on numerical solution of the corresponding Maxwell-Bloch equations. The analytical theory was not derived and the conditions for the pulse formation were not optimized, which resulted in moderate gain for the seeding XUV / X-ray field and moderate contrast of the produced sub-fs pulses, as well as in domination of the amplified spontaneous emission (ASE) from the active medium over the generated signal.

The present work is a logical continuation of [22]. Here we derive an analytical solution for the XUV radiation at the output of a hydrogen-like plasma-based X-ray laser seeded by a quasi-monochromatic resonant XUV field and simultaneously irradiated by a strong far-off-resonance infrared (IR) / optical field. The analytical theory allows us to qualitatively understand the influence of (i) the gain magnitude at the resonant XUV transition and (ii) the plasma dispersion at the frequency of the modulating IR field on the sub-fs pulse formation. The possibility for sub-fs pulse formation is shown for a wide range of the gain values at the XUV transition and for various wavelengths of the modulating field. The results of the analytical solution are verified by the numerical calculations, which account for a variety of nonlinear processes in the system under consideration. Based on both the analytical theory and extensive numerical studies, we find the optimal conditions for the sub-fs pulse formation in the active medium of $Li^{2+}$ plasma-based hydrogen-like X-ray laser [17]. We show the possibility to produce the pulse trains with the pulse duration of 400 -600 as and peak intensity in the range of $10^{10}$-$10^{11}$ W/cm$^2$ at the carrier wavelength 13.5 nm. This carrier wavelength corresponds to the maximum reflectivity of Si:Mo multilayer mirrors, which makes such pulses especially attractive for practical applications [24-27]. Under the optimal conditions, the influence of the amplified spontaneous XUV emission from the active medium is shown to be negligible. The effect of sub-fs pulse formation is robust and insensitive to moderate variations of the parameters of the active medium and the IR field and can be implemented under the experimentally feasible conditions.

The paper is organized as follows. In Sec. II we introduce the basic set of density matrix and wave equations, which describe the propagation of a resonant XUV field through the plasma of hydrogen-like ions in the presence of a strong modulating IR field. We also discuss a method to account for ASE, which emerges from the quantum noises of the medium and adversely affects the pulse formation by reducing the amplification of the seeding XUV field and overlapping with the produced attosecond pulse train in space and time. In Sec. III we derive an analytical solution for the output XUV field produced from the quasi-monochromatic incident radiation and reexamine the role of plasma dispersion in the pulse formation. On the basis of the analytical solution, we discuss the optimal conditions for the sub-fs pulse formation for various parameters of the active medium and the modulating field. In Sec. IV we present the results of numerical studies based on the general set of equations given in Sec. II. We verify the results of the linearized analytical theory taking into account the nonlinearities of the system and determine more precisely the optimal conditions for the sub-fs pulse formation feasible experimentally for the case of an active medium of $Li^{2+}$ hydrogen-like plasma-based X-ray laser. The results are summarized in Sec. V. Appendixes A and B present some technical derivations.

## II. THEORETICAL MODEL

Below we consider the amplification of a resonant seeding XUV radiation in the active medium of a recombination plasma-based X-ray laser with inversion at the transition $n=1 \leftrightarrow n=2$ of hydrogen-like ions (where $n$ is the principal quantum number) [17, 22, 23, 28, 29], dressed by a moderately strong optical laser field. We assume that at the entrance to the medium, $x=0$, the seeding XUV/X-ray field is quasi-monochromatic and linearly polarized:

$$\vec{E}(x=0,t) = \frac{1}{2}\vec{z}_0 \tilde{E}_{inc}(t) \exp(-i\omega_{inc}t) + \text{c.c.}, \tag{1}$$

where $\vec{z}_0$ is a unit polarization vector, c.c. stands for complex conjugation, $\omega_{inc}$ is the carrier frequency of the field, which is close to the frequency of the inverted transition, and $\tilde{E}_{inc}(t)$ is a slowly varying amplitude of the incident field. In the following, we consider the two types of the function $\tilde{E}_{inc}(t)$. For the analytical solution in Sec. III we assume that the seeding field (1) is turned on instantly at time $t=0$ and has a constant amplitude after that, while for the numerical calculations in Sec. V we use an incident field with a rectangular envelope smoothly turned on and off.

The medium is assumed to be simultaneously irradiated by a modulating IR field, which has the same polarization and propagates in the same direction as the incident XUV radiation (1):

$$\vec{E}_\Omega(x,t) = \vec{z}_0 E_M \cos\left[\Omega\left(t - \frac{n_{pl}}{c}x\right)\right]. \tag{2}$$

Here $E_M$ and $\Omega$ are the amplitude and angular frequency of the modulating field, respectively; $c$ is the speed of light in vacuum and $n_{pl} = \sqrt{1 - 4\pi N_e e^2/(m_e \Omega^2)}$ is the plasma refraction index at the frequency of the modulating field; $N_e$ is the concentration of free electrons, while $e$ and $m_e$ are charge and mass of an electron, respectively. Equation (2) assumes that modulating field is monochromatic and propagates through the medium with the phase velocity $c/n_{pl}$. These approximations are justified if (i) the duration of the modulating field significantly exceeds (a) the duration of the incident XUV field, and (b) the relaxation times of the active medium; (ii) the plasma is stationary and uniform, so that the concentration of free electrons is nearly constant in the interaction volume during the interaction time (some variations of $n_{pl}$ shall not affect the results if $n_{pl} \simeq 1$); and (iii) the modulating field is far-detuned from all the transitions connecting the populated states of the ions to each other and to the other states.

The relevant energy level scheme describing an interaction of the resonant XUV radiation (1) with hydrogen-like ions includes five states of the ions, namely, the ground state $|1\rangle = |1s\rangle$, which corresponds to the energy level $n=1$, and the excited states $|2\rangle = (|2s\rangle + |2p,m=0\rangle)/\sqrt{2}$, $|3\rangle = (|2s\rangle - |2p,m=0\rangle)/\sqrt{2}$, $|4\rangle = |2p,m=1\rangle$, and $|5\rangle = |2p,m=-1\rangle$, which correspond to the energy level $n=2$. Here $m$ is a projection of the orbital moment of the ions on the polarization direction of the modulating optical field (z-axis). Under the action of the modulating field (2), the upper resonant energy level of the ions is split into three sublevels due to the Stark effect. The two of them, corresponding to the states $|2\rangle$ and $|3\rangle$, oscillate in space and time along with the electric field strength of the modulating field due to the linear Stark effect and also experience a time-independent shift due to the quadratic Stark effect. The third sublevel corresponds to the states $|4\rangle$ and $|5\rangle$, which remain degenerate and experience only a quadratic Stark effect. The resonant polarization of the medium has a form

$$\vec{P}(x,t) = N_{ion}\left[\vec{d}_{12}\rho_{21} + \vec{d}_{13}\rho_{31} + \vec{d}_{14}\rho_{41} + \vec{d}_{15}\rho_{51} + \text{c.c.}\right], \tag{3}$$

where $N_{ion}$ is concentration of the resonant ions, $\vec{d}_{1i}$ are electric dipole moments of the transitions $|i\rangle \leftrightarrow |1\rangle$, $i=2,3,4,5$, while $\rho_{i1}$ are quantum coherencies at these transitions. The values of dipole moments $\vec{d}_{1i}$ are given by $\vec{d}_{12} = \vec{z}_0 d_{tr}$, $\vec{d}_{13} = -\vec{z}_0 d_{tr}$, $\vec{d}_{14} = \vec{d}_{15} = i\vec{y}_0 d_{tr}$, where $d_{tr} = \dfrac{2^7}{3^5 Z} ea_0$ and $a_0$ is the Bohr radius. The nonzero elements of the dipole moment matrix include also $\vec{d}_{22} = \vec{z}_0 d_{av}$ and $\vec{d}_{33} = -\vec{z}_0 d_{av}$, where $d_{av} = \dfrac{3}{Z} ea_0$. As discussed in [22], the incident z-polarized XUV field (1) is amplified and generates sidebands due to the interaction with the transitions $|2\rangle \leftrightarrow |1\rangle$ and $|3\rangle \leftrightarrow |1\rangle$, whose dipole moments are oriented along z-axis. The transitions $|4\rangle \leftrightarrow |1\rangle$ and $|5\rangle \leftrightarrow |1\rangle$, whose dipole moments are perpendicular to z-axis, serve as a source of ASE, which is polarized along y-axis and reduces the gain for the z-polarized XUV field by increasing ground-state population. In the following, we consider a plasma, which consists of only the resonant hydrogen-like ions and free electrons, so that the ion concentration is expressed through the electron concentration as $N_{ion} = N_e/(Z-1)$, where $Z$ is the ion nucleus charge number. The time evolution of the quantum state of the ions is described by the density-matrix equations:

$$\begin{cases} \dfrac{\partial \rho_{11}}{\partial t} = \gamma_{11} \sum_{k=2}^{5} \rho_{kk} - i\left[\hat{H}, \hat{\rho}\right]_{11}, \\ \dfrac{\partial \rho_{ij}}{\partial t} = -\gamma_{ij} \rho_{ij} - i\left[\hat{H}, \hat{\rho}\right]_{ij}, \ i,j = \{1,2,3,4,5\}, \ ij \neq 11, \end{cases} \quad (4)$$

which take into account that the ground state $|1\rangle$ does not decay into the other states. In Eqs. (4) $\gamma_{ij}$ are decay rates of the density matrix elements $\rho_{ij}$, while $\hat{H}$ is the Hamiltonian of the system. In the presence of both the resonant XUV field and the modulating optical field it has a matrix form

$$\hat{H} = \begin{pmatrix} \hbar\omega_1 & -E_z d_{tr} & E_z d_{tr} & -iE_y d_{tr} & -iE_y d_{tr} \\ -E_z d_{tr} & \hbar\omega_2(t,x) & 0 & 0 & 0 \\ E_z d_{tr} & 0 & \hbar\omega_3(t,x) & 0 & 0 \\ iE_y d_{tr} & 0 & 0 & \hbar\omega_4 & 0 \\ iE_y d_{tr} & 0 & 0 & 0 & \hbar\omega_5 \end{pmatrix}, \quad (5)$$

where

$$\begin{cases} \hbar\omega_1 = -\dfrac{m_e e^4 Z^2}{2\hbar^2}\left\{1 + \dfrac{9}{256}F_0^2\right\}, \\ \hbar\omega_2(t,x) = -\dfrac{m_e e^4 Z^2}{8\hbar^2}\left\{1 + \dfrac{21}{4}F_0^2 + 3F_0 \cos\left(\Omega\left[t - n_{pl}\dfrac{x}{c}\right]\right)\right\}, \\ \hbar\omega_3(t,x) = -\dfrac{m_e e^4 Z^2}{8\hbar^2}\left\{1 + \dfrac{21}{4}F_0^2 - 3F_0 \cos\left(\Omega\left[t - n_{pl}\dfrac{x}{c}\right]\right)\right\}, \\ \hbar\omega_4 = \hbar\omega_5 = -\dfrac{m_e e^4 Z^2}{8\hbar^2}\left\{1 + \dfrac{39}{8}F_0^2\right\} \end{cases} \quad (6)$$

are the energies of the states $|1\rangle$, $|2\rangle$, $|3\rangle$, $|4\rangle$, and $|5\rangle$, correspondingly, taking into account the linear and quadratic Stark effect induced by the modulating field (2) [30]; $F_0 = \left(\dfrac{2}{Z}\right)^3 \dfrac{E_M}{E_A}$ is a di-mensionless amplitude of the modulating field, and $E_A = m_e^2 e^5 / \hbar^4 \cong 5.14 \cdot 10^9 \, V/cm$ is the atomic unit for electric field. In turn, the decay rates $\gamma_{ij}$ are determined by the following equations:

$$\gamma_{21} = \gamma_{31} = \gamma_{coll} + w_{ion}^{(2,3)}/2 + \Gamma_{rad}/2 \equiv \gamma_z,$$

$$\gamma_{41} = \gamma_{51} = \gamma_{coll} + w_{ion}^{(4,5)}/2 + \Gamma_{rad}/2 \equiv \gamma_y,$$

$$\gamma_{32} = \gamma_{coll} + w_{ion}^{(2,3)} + \Gamma_{rad}, \; \gamma_{54} = \gamma_{coll} + w_{ion}^{(4,5)} + \Gamma_{rad}, \quad (7)$$

$$\gamma_{42} = \gamma_{52} = \gamma_{43} = \gamma_{53} = \gamma_{coll} + w_{ion}^{(2,3)}/2 + w_{ion}^{(4,5)}/2 + \Gamma_{rad},$$

$$\gamma_{11} = \Gamma_{rad}, \; \gamma_{22} = \gamma_{33} = w_{ion}^{(2,3)} + \Gamma_{rad}, \; \gamma_{44} = \gamma_{55} = w_{ion}^{(4,5)} + \Gamma_{rad}.$$

In (7) $\gamma_{coll}$ is the collisional broadening of the spectral lines, $\Gamma_{rad}$ is the radiative decay rate from each of the upper states of the ions, |2⟩, |3⟩, |4⟩, or |5⟩, to the ground state |1⟩, while $w_{ion}^{(2,3)}$ and $w_{ion}^{(4,5)}$ are the ionization rates from the states |2⟩ or |3⟩, and |4⟩ or |5⟩, respectively, averaged over the cycle of the modulating field:

$$w_{ion}^{(2,3)} = \frac{m_e e^4 Z^2}{16 \hbar^3} \sqrt{\frac{3 F_0}{\pi}} \left[ \frac{4}{F_0} e^3 + \left(\frac{4}{F_0}\right)^3 e^{-3} \right] e^{-\frac{2}{3 F_0}},$$

$$w_{ion}^{(4,5)} = \frac{m_e e^4 Z^2}{4 \hbar^3} \sqrt{\frac{3 F_0}{\pi}} \left(\frac{4}{F_0}\right)^2 e^{-\frac{2}{3 F_0}} \quad (8)$$

Since the modulating field should not ionize the active medium during the interaction time, it is assumed to be not too strong so that both (i) the quadratic Stark shifts of the resonant energy levels and (ii) the ionization rates from them are much smaller than the frequency of the modulating field, $\Omega$. It allows taking them into account as time-independent values [31, 32], as it is done in (6) and (8). Under the same conditions the cubic Stark effect and higher-order corrections to the Stark shift can be safely neglected.

In the following, we assume the active medium in the form of a long cylinder of the length $L$ and radius $R \ll L$, with Fresnel parameter, $F = \pi R^2 / (\lambda_{21} L)$, on the order of unity (here $\lambda_{21} = 2\pi c / \bar{\omega}_{21}$ is the wavelength of the resonant XUV field, and $\bar{\omega}_{21}$ is the time-averaged frequency of the transition |2⟩↔|1⟩), and describe the propagation of the XUV radiation through the medium by one-dimensional wave equation:

$$\frac{\partial^2 \vec{E}}{\partial x^2} - \frac{\varepsilon_{X-ray}}{c^2} \frac{\partial^2 \vec{E}}{\partial t^2} = \frac{4\pi}{c^2} \frac{\partial^2 \vec{P}}{\partial t^2}, \quad (9)$$

where $\vec{E} = \vec{z}_0 E_z + \vec{y}_0 E_y$ is a vector of the resonant field inside the medium (although the seeding field (1) is z-polarized, the y-polarized component appears in the medium because of the spontaneous emission at the transitions |4⟩↔|1⟩ and |5⟩↔|1⟩), $\vec{P}$ is a vector of the resonant polarization (3), and $\varepsilon_{X-ray} = 1 - 4\pi N_e e^2 / (m_e \omega_{inc}^2) \simeq 1$ is a dielectric permittivity of the plasma for the XUV field. Equations (3)-(9) describe the transformation of the resonant XUV field (1) in the active medium of a hydrogen-like plasma-based X-ray laser irradiated by the modulating field (2). We further introduce a change of independent variables, $x, t \rightarrow x, \tau = t - x\sqrt{\varepsilon_{X-ray}}/c$, and seek for a solution of this system within the slowly varying amplitude approximation for both the XUV field and the resonant polarization, and the rotating wave approximation for the density matrix elements as follows:

$$\begin{cases} \vec{E}(x,\tau) = \dfrac{1}{2}\{\vec{z}_0 \tilde{E}_z(x,\tau) + \vec{y}_0 \tilde{E}_y(x,\tau)\} e^{-i\omega_{inc}\tau} + \text{c.c.}, \\ \vec{P}(x,\tau) = \dfrac{1}{2}\{\vec{z}_0 \tilde{P}_z(x,\tau) + \vec{y}_0 \tilde{P}_y(x,\tau)\} e^{-i\omega_{inc}\tau} + \text{c.c.}, \end{cases} \quad (10a)$$

$$\begin{cases} \rho_{i1}(x,\tau) = \tilde{\rho}_{i1}(x,\tau) e^{-i\omega_{inc}\tau}, \; i = \{2,3,4,5\}, \\ \rho_{ij}(x,\tau) = \tilde{\rho}_{ij}(x,\tau), \; ij \neq \{21,31,41,51\}, \end{cases} \quad (10b)$$

where $\tilde{E}_p$, $\tilde{P}_p$ ($p = z, y$) and $\tilde{\rho}_{ij}$ ($i,j = 1 \div 5$) are slowly varying functions of space and time, which means $\dfrac{1}{|\tilde{E}_p|}\left|\dfrac{\partial \tilde{E}_p}{\partial x}\right|, \dfrac{1}{|\tilde{P}_p|}\left|\dfrac{\partial \tilde{P}_p}{\partial x}\right|, \dfrac{1}{|\tilde{\rho}_{ij}|}\left|\dfrac{\partial \tilde{\rho}_{ij}}{\partial x}\right| \ll \omega_{inc}\sqrt{\varepsilon_{X-ray}}/c$ and $\dfrac{1}{|\tilde{E}_p|}\left|\dfrac{\partial \tilde{E}_p}{\partial \tau}\right|, \dfrac{1}{|\tilde{P}_p|}\left|\dfrac{\partial \tilde{P}_p}{\partial \tau}\right|, \dfrac{1}{|\tilde{\rho}_{ij}|}\left|\dfrac{\partial \tilde{\rho}_{ij}}{\partial \tau}\right| \ll \omega_{inc}$. Then, Eqs. (9) and (4) take the form

$$\begin{cases} \dfrac{\partial \tilde{E}_z}{\partial x} = i \dfrac{4\pi \omega_{inc} N_{ion} d_{tr}}{c\sqrt{\varepsilon_{X-ray}}} (\tilde{\rho}_{21} - \tilde{\rho}_{31}), \\ \dfrac{\partial \tilde{E}_y}{\partial x} = -\dfrac{4\pi \omega_{inc} N_{ion} d_{tr}}{c\sqrt{\varepsilon_{X-ray}}} (\tilde{\rho}_{41} + \tilde{\rho}_{51}). \end{cases} \quad (11)$$

The explicit form of the density matrix equations (4)-(6) under the considered approximations (10) is given in [22] and will not be reproduced here because of its bulkiness. To solve the resulting system of equations (4)-(6) and (11), one should set (i) the boundary conditions for the resonant field at the entrance of the medium, and (ii) the initial conditions for the density matrix elements.

Since $\varepsilon_{X-ray} \simeq 1$, the reflection of the incident XUV field from the edges of the medium is negligible. Thus, the boundary conditions for the resonant field are

$$\tilde{E}_z(x=0,\tau) = \tilde{E}_{inc}(\tau), \; \tilde{E}_y(x=0,\tau) = 0. \quad (12)$$

Similarly to [29], as initial conditions we assume that at $\tau = 0$ the ions are excited to the states |2⟩-|5⟩ with equal probability by a running wave of a pumping laser field. Thus, we have

$$\tilde{\rho}_{11}(x,\tau=0) = \dfrac{1-4n_{tr}}{5}, \text{ and } \tilde{\rho}_{ii}(x,\tau=0) = \dfrac{1+n_{tr}}{5} \text{ for } i = \{2,3,4,5\}, \quad (13)$$

where $n_{tr}$ is the initial population difference at the transitions |i⟩↔|1⟩, $i = \{2,3,4,5\}$. However, contrary to [22], we assume random initial values of the quantum coherencies at the inverted transitions, $\tilde{\rho}_{i1}(x,\tau=0)$, which we use to take into account the spontaneous emission at these transitions [33]. The initial values of the quantum coherencies are calculated in Appendix A.

The resulting system of equations is rather difficult to analyze. Thus, in order to get an insight into the process of sub-fs pulse formation from the quasi-monochromatic seeding XUV radiation, in the next Section we derive a simplified analytical solution, which is further compared to the results of numerical calculations based on equations (4)-(6), and (11) with boundary conditions (12) and initial conditions for the density matrix elements given by Eq. (13) and Appendix A.

## III. ANALYTICAL STUDY

In this section, we will assume that the incident z-polarized XUV field (1) is exactly resonant to the transitions |2⟩↔|1⟩ and |3⟩↔|1⟩ (with dipole moments oriented along z-axis) taking into account the quadratic Stark shift of frequencies of these transitions by the modulating IR field. In such a case, the carrier frequency of the seeding field (1) is

$$\omega_{inc} = \bar{\omega}_{21} = \bar{\omega}_{31} = \frac{3m_e e^4 Z^2}{8\hbar^3}\left(1 - \frac{109}{64}F_0^2\right) \equiv \omega_z. \tag{14}$$

According to Eqs. (11), the transformation of z-polarized XUV / X-ray field is governed by the slowly varying amplitudes of the coherencies $\tilde{\rho}_{21}$ and $\tilde{\rho}_{31}$, while the development of y-polarized XUV / X-ray field from spontaneous emission of the active medium is determined by the values of $\tilde{\rho}_{41}$ and $\tilde{\rho}_{51}$. In the linear approximation, assuming that the population differences between the resonant states of the ions are constant, $\tilde{\rho}_{ii} - \tilde{\rho}_{11} = n_{tr}$ for $i = \{2,3,4,5\}$, while the coherencies between the excited states of the ions are negligible, $\tilde{\rho}_{ij} = 0, i \neq j, i, j \neq 1$, the values $\tilde{\rho}_{21}$, $\tilde{\rho}_{31}$, and $\tilde{\rho}_{41}$, $\tilde{\rho}_{51}$ evolve independently of each other. This is correct at the initial stage of evolution of the active medium, until the coupling between the resonant XUV field and the medium results in redistribution of the population of states and excitation of coherencies at the transitions $|i⟩↔|j⟩$, $i, j \neq 1$. In such a regime, z- and y-polarization components of the XUV radiation are amplified independently of each other, and their propagation through the medium can be analyzed separately.

We start by making a comment on the evolution of the y-polarized field. According to Eqs. (11), the intensity of y-polarized field at the very early stage of its development can be estimated as $I_y(\tau = 0) \sim \langle|\tilde{\rho}_{41} + \tilde{\rho}_{51}|^2\rangle = [\text{at } \tau = 0] = \langle|\tilde{\rho}_{41}|^2\rangle + \langle|\tilde{\rho}_{51}|^2\rangle$. As follows from Eq. (A7) (see Appendix A), $\langle|\tilde{\rho}_{i1}(\tau = 0)|^2\rangle \sim \langle|A_{i,k}|^2\rangle/R^4$ and $\langle|A_{i,k}|^2\rangle = N_{ik} \sim R^2$, so that $I_y(\tau = 0) \sim 1/R^2$, where $R$ is the radius of the active medium (plasma channel). Thus, an intensity of the spontaneously emitted y-polarized field is reduced with increasing radius of the plasma channel.

We further consider z-polarized XUV field. Since the resonant ions possess inversion symmetry with respect to the change of sign of the modulating optical field, under the action of the monochromatic field (2) their steady-state response to the resonant XUV radiation (1) should be a periodic function of time with the period equal to $\pi/\Omega$, and the output resonant radiation should have the same periodicity. Thus, it is natural to search for a nonstationary solution for the output XUV field in the form

$$\tilde{E}_z(x,\tau) = \sum_{l=-\infty}^{\infty} E_l(x,\tau) e^{-i2l\Omega\tau}, \tag{15}$$

assuming that the amplitudes of its spectral components are slowly varying functions of space and time. Moreover, we will further assume that the central (resonant) spectral component of the slowly varying amplitude of z-polarized field dominates the sidebands at any time and any position in space:

$$|E_0(x,\tau)| \gg |E_l(x,\tau)| \quad (\forall l, x, \tau). \tag{16}$$

As we will see below from the comparison of the analytical solution with the results of numerical calculations, this is a sufficiently good approximation for the Li$^{2+}$ X-ray lasers [17, 28, 29]. We will also assume that the incident z-polarized XUV field (1) is sufficiently strong, so that z-polarized spontaneous emission can be neglected and $\tilde{\rho}_{21}(x, \tau = 0) = \tilde{\rho}_{31}(x, \tau = 0) = 0$. To derive a simple analytical solution, we assume that the incident field (1) is turned on instantly at time $\tau|_{x=0} = t = 0$ and has thereafter a constant amplitude:

$$\tilde{E}_z(x=0,\tau) = \theta(\tau)E_{inc}, \tag{17}$$

where $\theta(\tau)$ is Heaviside step function: $\theta(\tau) = 0$ for $\tau < 0$ and $\theta(\tau) = 1$ for $\tau \geq 0$. Within the described approximations, the evolution of z-polarized XUV/X-ray field during its propagation through the medium is governed by the equations

$$\begin{cases} \dfrac{\partial \tilde{E}_z}{\partial x} = i \dfrac{4\pi\omega_z N d_{tr}}{c\sqrt{\varepsilon_{X-ray}}} (\tilde{\rho}_{21} - \tilde{\rho}_{31}), \\ \dfrac{\partial \tilde{\rho}_{21}}{\partial \tau} + \left( -i\Delta_\Omega \cos\left(\Omega\tau + (1-n_{pl})\dfrac{\Omega}{c}x\right) + \gamma_z \right)\tilde{\rho}_{21} = -\dfrac{i}{2\hbar} n_{tr} d_{tr} E_0(x,\tau), \\ \dfrac{\partial \tilde{\rho}_{31}}{\partial \tau} + \left( i\Delta_\Omega \cos\left(\Omega\tau + (1-n_{pl})\dfrac{\Omega}{c}x\right) + \gamma_z \right)\tilde{\rho}_{31} = \dfrac{i}{2\hbar} n_{tr} d_{tr} E_0(x,\tau). \end{cases} \tag{18}$$

Here $\Delta_\Omega = \dfrac{3m_e e^4 Z^2}{8\hbar^2} F_0$ is the depth of modulation of frequencies of the transitions $|2\rangle\leftrightarrow|1\rangle$ and $|3\rangle\leftrightarrow|1\rangle$ due to the linear Stark effect. As shown in Appendix B, the amplitudes of the spectral components of z-polarized field (15), which satisfy equations (18) and boundary condition (17), are:

$$E_0(x,\tau) = E_{inc} \exp\{g(P_\Omega,\tau)x\}, \tag{19a}$$

$$E_l(x,\tau) = E_{inc} \dfrac{J_{2l}(P_\Omega)}{J_0(P_\Omega)} g(P_\Omega,\tau) \dfrac{\exp\{(g(P_\Omega,\tau) - i2l\Delta K)x\} - 1}{g(P_\Omega,\tau) - i2l\Delta K}, \, l \neq 0, \tag{19b}$$

$$g(P_\Omega,\tau) = g_0 J_0^2(P_\Omega)(1 - e^{-\gamma_z \tau}), \tag{19c}$$

where $g_0 = \dfrac{4\pi\omega_z n_{tr} d_{tr}^2 N_{ion}}{\hbar c \sqrt{\varepsilon_{X-ray}} \gamma_z}$ is the gain coefficient for the resonant XUV field without the modulation and $\Delta K = (1-n_{pl})\dfrac{\Omega}{c}$ is a change of the wavenumber of the modulating field due to the plasma dispersion. As follows from Eq. (19c), the effective gain, $g(P_\Omega,\tau)$, for the resonant spectral component of the XUV / X-ray field in the presence of the modulating field depends on time and the value of modulation index. The effective gain tends to zero for $\tau \ll \gamma_z^{-1}$ because of the finite response time of the ions. The gain is also zero for the values of modulation index satisfying the condition $J_0(P_\Omega) = 0$, in particular, for $P_\Omega \approx 2.4$, due to suppression of the resonant spectral component of the multifrequency polarization of the medium (see, for example, [34]). In turn, the sidebands (19b) are produced due to coherent scattering of the resonant spectral component (19a) on the ions with space-time-dependent transition frequencies and strongly depend on the relative influence of (i) the plasma dispersion on the modulating IR field and (ii) the amplification for the resonant spectral component of the XUV radiation.

In the following, we consider an active medium consisting of $Li^{2+}$ ions and free electrons with the density of the ions $N_{ion}=1.5\times10^{17}$ cm$^{-3}$ and the electron density $N_e=3\times10^{17}$ cm$^{-3}$. The ion temperature is assumed to be 1 eV, and the electron temperature is 2 eV. These parameters are the same as in [22] and close to those in the experiment [17]. For such plasma parameters, the collisional and radiative decay times are $\gamma_{coll}^{-1} \approx 0.425\,\text{ps}$ [35] and $\Gamma_{rad}^{-1} \approx 19.7\,\text{ps}$, respectively. We also assume the same intensity of the modulating field as in [22], $I_{IR}=4\times 10^{14}$ W/cm$^2$, which is slightly below the threshold for rapid tunnel ionization of the active medium. Such modulation intensity assumed throughout the paper corresponds to the ionization rates $1/w_{ion}^{(2,3)} = 3.3\,\text{ps}$. In such a case, $\gamma_z^{-1} = 395\,\text{fs}$. Similarly to [22], we also assume that all the ions are initially in excited states and $n_{tr} = 1/4$.

In Figs. 1-3 we plot the dependencies of the amplitudes of 0th, ±1st and ±2nd spectral components of the amplified XUV field on the propagation distance, calculated according to the analytical solution (15) and (19), for the above parameters of the active medium and different moments of local time: $\gamma_z \tau = 0.01$ in Fig. 1, $\gamma_z \tau = 0.2$ in Fig. 2 and $\gamma_z \tau = 3$ in Fig. 3. In all cases, the value of modulation index is $P_\Omega = 3.3$, which corresponds to the wavelength of the modulating field in vacuum $\lambda_{IR} = 2\pi c/\Omega \approx 1400\,\text{nm}$. For this modulating field, the above parameters of the active medium give $g_0/\Delta K = 13.8$. As follows from Eq. (19c), for the very initial moments of time, $\tau \ll \gamma^{-1}$, the effective gain coefficient is small, so that the condition

$$g(P_\Omega,\tau) \ll |l|\Delta K, \tag{20}$$

is satisfied for arbitrary $l$. The condition (20) can be fulfilled also if one of the following statements is true: (i) the unperturbed gain coefficient, $g_0$, is small, (ii) the number of a sideband is large, $|l| \gg 1$, or (iii) the modulation index satisfies the condition $J_0(P_\Omega) \approx 0$. In the case (20) the amplitude of $l$-th sideband takes the form

$$E_l(x,\tau) = E_{inc} \frac{J_{2l}(P_\Omega)}{J_0(P_\Omega)} \frac{g(P_\Omega,\tau)}{l\Delta K} \sin(l\Delta K x)\exp\{-il\Delta K x\}. \tag{21}$$

As follows from Eq. (21), in the case of a weak gain / strong plasma dispersion (i) the amplitudes of all the sidebands are small compared to the central (resonant) spectral component of the field, whereas the amplitude of each sideband (ii) decreases with increasing sideband number $\sim 1/|l|$ and (iii) oscillates in space reaching the maximum values at the so-called coherence length $L_{coh}^{(l)} = \pi/(2\Delta K|l|)$, which is inversely proportional to the sideband number and can be written as

$$L_{coh}^{(l)} = \frac{\lambda_{IR}}{4|l|(1-n_{pl})}. \tag{22}$$

The oscillatory nature of the amplitudes of sidebands of the XUV field and their decrease with increasing sideband number originate from interference of the contributions to the sidebands (coherently scattered fields) generated at different depths of the medium. Under the action of the modulating field, the energies of the resonant states of the ions and the corresponding transition frequencies vary in space and time in accordance with Eqs. (6), i.e., follow a traveling wave of modulation, which propagates through the medium with the phase velocity of the modulating field, $c/n_{pl}$. The sidebands appear due to coherent scattering of the resonant spectral component of the XUV field on this modulation wave. But since the phase velocity of the modulation differs from the phase velocity of the XUV field, $c/\sqrt{\varepsilon_{X-ray}}$, at each depth of the medium, $x$, the sidebands are generated with a phase shift with respect to the resonant spectral component, which is equal to $2l\Delta K x$. The amplitude of a sideband at the output of the medium results from a sum of the contributions, which were generated at different depths of the medium, and thus possess different phase shifts with the respect to the resonant spectral component. At the coherence length, constructive interference of these contributions results in a maximum amplitude of the sideband, whereas at the doubled coherence length destructive interference of the contributions originating from different depth of the medium reduces the sideband amplitude to the minimum. If the amplification of the resonant spectral component at the coherence length of a sideband is not substantial, as implied by (20) and as it is in Fig. 1, then at even multiples of the coherence length the amplitude of a sideband tends to zero. However, as soon as the amplification becomes considerable, the contributions to a sideband generated at subsequent odd and even multiples of the coherence length do not compensate each other anymore, so that the oscillations of a sideband amplitude become (i) less pronounced and (ii) accompanied by its exponential grows, see Fig. 2.

With increasing time, the output field tends to the stationary solution, which has the same form as (19), except for the change in the effective gain coefficient:

$$g(P_\Omega,\tau) \to g_{St}(P_\Omega) = g_0 J_0^2(P_\Omega) \tag{23}$$

and is valid for $\gamma_z \tau \gg 1$. Such a case is represented in Fig. 3, where solid lines correspond to the parameter values of Figs. 1 and 2, while dashed lines are plotted for the two times smaller gain coefficient, $g_0/\Delta K = 6.9$. As we can see, in the case of $g_0/\Delta K = 13.8$ the steady-state analytical solution runs out of its range of applicability, since the amplitudes of ±1 sidebands reach the amplitude of the resonant spectral component and the condition (16) fails. At the same time, for $g_0/\Delta K = 6.9$ the amplitudes of sidebands are approximately two times smaller, so that the solution (19) and (23) remains valid at least qualitatively. In the steady-state regime, $\tau \gg \gamma_z^{-1}$, the relative amplitudes of sidebands reach their maximum values, while their spatial oscillations almost fully disappear due to the strong amplification of the field. It is also worth noting that for weaker gain (smaller $g_0$), or stronger plasma dispersion (larger $\Delta K$), the steady-state solution predicts exactly the same spatial dependencies of the amplitudes of sidebands as those given by the solution (19) for smaller local times, $\tau$. Thus, Fig. 2 is reproduced by the stationary solution for $g_0/\Delta K = 2.5$.

From the analytical solution one can make the following conclusions. (i) The sidebands are intensified with increasing gain for the resonant XUV radiation and/or reducing plasma dispersion for the modulating optical field. (ii) Within the range of applicability of the analytical solution given by Eq. (16), only the resonant spectral component and ±1 sidebands may have comparable amplitudes. (iii) These spectral components of the XUV field are always phase-aligned, that is, their phases lie on the same line, while the parameters of the problem determine only its incline. Thus, if the amplification is sufficiently strong and the modulation index is chosen in such a way that the amplitudes of ±1 sidebands are comparable to the amplitude of the resonant component, the output XUV field will acquire a form a sub-fs pulse train. Within the range of applicability of the analytical solution, the contrast of the produced pulses is limited by a relatively narrow bandwidth (only three spectral components) of the output field and domination of its central spectral component over the sidebands. However, since in the limit of a strong gain the condition (16) is not satisfied and the analytical solution is not valid, we further proceed with the study of the optimal conditions for sub-fs pulse formation on the basis of numerical solution of Eqs. (4)-(6) and (11) under the rotating wave approximation (RWA) and slowly varying envelope (SVE) approximation. As shown below, the pulses with better characteristics (shorter duration, higher intensity and contrast) can be produced in a strong gain limit.

## IV. PULSE FORMATION IN HYDROGEN-LIKE PLASMA OF LI$^{2+}$

In this section, we will analyze the optimal conditions for sub-fs pulse formation from the XUV radiation on the basis of numerical solution of Eqs. (4)-(6) and (11) taking into account the rescattering of sidebands into each other, their nonlinear interaction with the medium (in particular, reduction of the population differences at the resonant transitions), and the amplified sponta-neous emission of y-polarization. In order to directly compare the results of the calculations with the analytical solution derived in the previous section, we assume an incident XUV field with a rectangular shape and smoothed turn-on and turn-off:

$$\tilde{E}_{inc}(t) = E_0 \times \begin{cases} \sin^2\left(\frac{\pi}{2}\frac{t}{t_{switch}}\right), & 0 \leq t < t_{switch}, \\ 1, & t_{switch} \leq t < t_{flat} + t_{switch}, \\ \cos^2\left(\frac{\pi}{2}\frac{[t-\{t_{flat}+t_{switch}\}]}{t_{switch}}\right), & t_{flat}+t_{switch} \leq t < t_{flat}+2t_{switch}, \\ 0, & t \geq t_{flat}+2t_{switch}, \end{cases} \quad (24)$$

where 1.5 ps $\leq t_{flat} \leq$ 3ps, depending on the parameters of the problem, and $t_{switch} = 15\,fs$, which is much larger than the period of the resonant XUV field, $2\pi/\omega_{inc} = 45\,as$. The radius of the plasma channel assumed to be $R=2.5$ μm.

As mentioned above, in a medium with a sufficiently high gain, z-polarized XUV radiation might acquire a form of a sub-fs pulse train. In Fig. 4(a) we plot the contrast of the produced pulses vs. the propagation distance through the medium, $x$, and the modulation index, $P_\Omega$. The pulse contrast, $C$, is defined as a difference between the maximum and minimum values of intensity of z-polarized XUV radiation within a half-cycle of the modulating field, normalized to the mean value of this intensity, averaged over the same time interval:

$$C = \left(\max\{I_z(x,\tau)\} - \min\{I_z(x,\tau)\}\right)/\mathrm{mean}\{I_z(x,\tau)\}, \quad (25)$$

where $I_z = \frac{c}{8\pi}|\tilde{E}_z(x,\tau)|^2$. For every $x$ and $P_\Omega$ Fig. 4(a) shows the pulse contrast at the peak of the generated pulse train (for generally different values of $\tau$). The pulse contrast reaches the maximum values C ≈ 3.5 in two regions: (i) 0.9 mm $\leq x \leq$ 1.4mm, 2.1 $\leq P_\Omega \leq$ 2.3 (which corresponds to 0.23π $\leq \Delta Kx \leq$ 0.36π), and (ii) 2.3 mm $\leq x \leq$ 3.3 mm, 3.2 $\leq P_\Omega \leq$ 3.4 (which corresponds to 0.87π $\leq \Delta Kx \leq$ 1.25π). As mentioned before, the intensity of the modulating field is fixed to $I_{IR}=4\times10^{14}$ W/cm$^2$. The modulation index is changed by changing the frequency of the modulating field, $\Omega$, which changes also its phase incursion acquired due to plasma dispersion. Thus, the same propagation distances for different $P_\Omega$ correspond to different values of $\Delta Kx$. The dependence of the pulse contrast on the modulation index in Fig. 4(a) is nearly the same as that predicted by the analytical theory and can be understood in the following way. At zero modulation index, there are no sidebands and the contrast is zero. With increasing $P_\Omega$ the amplitudes of sidebands grow, which leads to the increasing pulse contrast, until the modulation index reaches the values $P_\Omega \approx 2.4$, at which $J_0(P_\Omega) \approx 0$ and the resonant interaction of z-polarized XUV field with the medium is suppressed (in particular, $g(P_\Omega,\tau) \approx 0$ for any $\tau$), so that the seeding field (1) traverses the medium with minimum modifications. With further increase in the modulation index the amplitudes of sidebands grow up again and the pulse contrast reaches its global maximum at $P_\Omega \approx 3.3$, near the maximum of the absolute value of the product $J_0(P_\Omega)J_2(P_\Omega)$, where the amplitudes of ±1 sidebands (19b) reach their maximum value. A subsequent increase in $P_\Omega$ reduces the pulse contrast because of the decrease of the relative amplitudes of ±1 sidebands. The dependence of the pulse contrast on the propagation distance is more complicated and cannot be understood without taking into account rescattering of the sidebands and nonlinear interaction between the resonant XUV field and the medium. While within the range of applicability of the analytical solution the maximum pulse contrast is achieved at $\Delta Kx \approx \pi/2$, which corres-ponds to the coherence length of ±1 sidebands of the XUV field, in Fig. 4(a) the optimal values of the propagation distance are either considerably smaller, $\Delta Kx \approx \pi/4$-$\pi/3$ for $P_\Omega \approx 2.2$, or considerably larger, $\Delta Kx \approx \pi$ for $P_\Omega \approx 3.3$.

In Fig. 4(b) we plot the ratio between the peak intensities of z-polarized and y-polarized components of the output XUV field (which correspond to the sub-fs pulse train and the ASE, respectively) in the logarithmic scale as a function of $x$ and $P_\Omega$. With increasing propagation distance, the relative intensity of the ASE grows because of the stronger amplification of y-polarized radiation. Indeed, since the transitions $|4\rangle \leftrightarrow |1\rangle$ and $|5\rangle \leftrightarrow |1\rangle$ are unaffected by the modulating field, except for a slight time-independent quadratic Stark shift of their frequencies, the gain coefficient for the y-polarized field is equal to its unperturbed value, $g_0$, which is always larger than that for the z-polarized field (19c), (23). Thus, in order to keep the ASE weak with respect to the generated pulse train, one should preferably use not too thick active medium.

In Figs. 5(a) we show the time dependence of intensity of the output XUV field for $x$=1.1 mm and $P_\Omega$=2.2, which correspond to the left optimal region of the parameter values in Fig. 4(a). The wavelength of the modulating field is $\lambda_{IR}$=930 nm, while its intensity is $I_{IR}$=4×10$^{14}$ W/cm$^2$. The peak intensity of the seeding field (1) is chosen to be $I_{inc}$=10$^9$ W/cm$^2$, which can be achieved by a similar unmodulated X-ray laser. As seen from Fig. 5(a), the peak intensity of the produced pulse train is approximately 20 times higher than the intensity of the incident XUV radiation, the duration of pulses is 440 as, while the pulse repetition period is 1.55 fs, which is a half-period of the modulating field. In Fig. 5(b) we show the windowed spectrum of the output XUV radiation, using a time window of the form $\sin^2(\pi(\tau-\tau_0)/\tau_{wind})$ with the full duration of 10 cycles of the IR field, that is $\tau_{wind}$= 31 fs. Such a time window allows us to extract the spectrum of the pulse train in the vicinity of its maximum, which differs from the total spectrum of the XUV radiation because of the gradual grow of the amplitudes of sidebands in time. As follows from Fig. 5(b), the amplitudes of ±1 sidebands almost reach the amplitude of the resonant spectral component of the field, the spectral components are phase-aligned, and the produced pulses are nearly bandwidth-limited. A similar dependencies are shown in Figs. 6(a) and 6(b) for $x$=2.5 mm and $P_\Omega$=3.3, which correspond to the right optimal region of the parameter values in Fig. 4(a). The corresponding parameters of the modulating field are: $\lambda_{IR}$=1400 nm and $I_{IR}$=4×10$^{14}$ W/cm$^2$. Due to the larger thickness of the medium, the incident XUV field of the same intensity as in Fig. 5 is amplified much more efficiently (by approximately 500 times), so that the peak intensity of the pulse train reaches 5×10$^{11}$ W/cm$^2$. The pulse duration is 590 as and the pulse repetition period equals 2.33 fs. Due to the larger value of the modulation index, in Fig. 6(b) the amplitudes of sidebands are larger than in Fig. 5(b): the ±1 sidebands are stronger than the resonant spectral component and the ±2 sidebands become substantial, so that the resulting pulse contrast is higher. On the other hand, for the parameters of Fig. 5 the ratio between the peak intensities of z-polarized and y-polarized XUV fields is higher than that in Fig. 6: 1600/1 instead of 30/1. Thus, with less intense seeding field the parameters of Fig. 5 will be preferable for the pulse formation in order to keep the ASE small compared to the amplified signal. It is worth noting here that while Figs. 5(b) and 6(b) show the time-windowed spectra of the XUV radiation, the real spectrum of the output XUV field integrated over the whole duration of the generated attosecond pulse train corresponds to a set of much narrower spectral lines (with the relative bandwidth on the order of 10$^{-4}$-10$^{-5}$), which might be useful for the energy-resolved interferometric measurements, in particular, based on RABBIT technique [36].

At this point, let us compare the results of the present paper with those obtained in [22], where possibility in principle to produce a sub-fs pulse train via an optical modulation of a hydrogen-like plasma based X-ray laser was shown. In that work the modulation index was assumed to be $P_\Omega$=4.45, while the length of Li$^{2+}$ active medium was $x$=1.25 mm. For such parameter values the pulse contrast equals 2.1, while the pulse duration and the pulse repetition period are 960 as and 3.14 fs, respectively. The use of a seeding field with intensity $I_{inc}$=10$^9$ W/cm$^2$ results in the peak intensity of the pulse train 3.45×10$^{10}$ W/cm$^2$. Thus, in the present paper we have shown the possibility to reduce the pulse duration more than two times, to increase the pulse contrast by approximately 70%, and to produce the pulses with an order of magnitude higher peak intensity. Furthermore, a proper choice of the parameters of the medium has allowed us to eliminate the effect of the ASE of the orthogonal y-polarization.

Finally, in Fig. 7 we compare the results of the analytical solution, Eqs. (19a)-(19c), with numerical calculations based on Eqs. (4)-(6) and (11), for the time dependence of the output XUV intensity within the range of applicability of the analytical theory (for the two times smaller gain coefficient than that used in Figs. 4-6). Fig. 7 is plotted for $x$=1.15 mm and $P_\Omega$=3.3, which maximize the pulse contrast in the case of the reduced gain coefficient. As follows from this figure, the analytics is in a good agreement with the numerical solution until the peak of the envelope the generated pulse train, where the population differences at the inverted transitions are considerably changed by stimulated emission. A comparison of Figs. 7 and 6 (which correspond to the same value of the modulation index and the same repetition period of the produced pulses) confirms that: (i) the pulses can be produced in a wide range of the gain coefficients and (ii) the characteristics of the pulses are improved with increasing gain of the active medium.

## V. CONCLUSION

In this paper, we derived an analytical solution describing the process of attosecond pulse formation in active medium of hydrogen-like plasma based X-ray laser, modulated by an IR laser filed and seeded with the quasi-monochromatic XUV radiation. On the basis of the derived solution we studied the qualitative dependencies of characteristics of the produced pulses on the parameters of the active medium and the modulating field for the case of $Li^{2+}$ hydrogen-like plasma based X-ray laser. The analytical solution shows that under conditions of a moderate gain for the typical parameter values of the active medium of $Li^{2+}$ X-ray laser, only the ±1 sidebands of the XUV field, separated from the resonance by ±2 frequencies of the modulating field, can be of comparable amplitudes with its resonant spectral component. These sidebands are always phase-aligned with the resonant spectral component of the field, so that the optimization of the pulse formation predominantly corresponds to the maximization of their amplitudes. At the same time, in active medium with a high gain the output XUV spectrum is broader than that predicted by the analytical theory. An optimization performed on the basis of an extensive numerical study of the problem shows the possibility to produce the pulses with duration down to 440 as and peak intensity up to $5\times10^{11}$ W/cm$^2$ using a seeding field with intensity $10^9$ W/cm$^2$ at wavelength 13.5 nm. The generated pulses are bandwidth-limited. The influence of the amplified spontaneous emission under the optimal conditions is negligible. An experimental implementation of this concept might allow to construct a bright source of spectral combs and sub-fs pulse trains in the XUV range, near the peak of reflectivity of Si:Mo X-ray optics, suitable for the spectroscopic applications and time-resolved studies of the ultrafast optical process.

## VI. ACKNOWLEDGMENTS

We acknowledge support from Russian Foundation for Basic Research (RFBR, Grant No.18-02-00924), National Science Foundation (NSF, Grant No. PHY-150-64-67), the Robert A. Welch Foundation (Grant No. A-1261), as well as from AFOSR and ONR. The analytical studies presented in Sections II-IV were supported by the Ministry of Education and Science of the Russian Federation under contract No.14.W03.31.0032. V.A.A. acknowledges support by the Foundation for the Advancement of Theoretical Physics and Mathematics BASIS.

*Corresponding author: khairulinir@ipfran.ru

## APPENDIX A. THE INITIAL VALUES OF THE QUANTUM COHERENCIES

In accordance with Eq. (3), the slowly varying amplitude of the resonant polarization of the medium consists of four parts:

$$\begin{cases} \tilde{P}_z(x,\tau=0) = \tilde{P}_{21}(x,\tau=0) + \tilde{P}_{31}(x,\tau=0), \\ \tilde{P}_y(x,\tau=0) = \tilde{P}_{41}(x,\tau=0) + \tilde{P}_{51}(x,\tau=0), \end{cases} \tag{A1}$$

where

$$\tilde{P}_{i1}(x,\tau=0) = 2\exp\{i\phi_i\} N_{ion} d_{tr} \tilde{\rho}_{i1}(x,\tau=0), \tag{A2}$$
$$\phi_2 = 0, \quad \phi_3 = \pi, \quad \phi_4 = \phi_5 = \pi/2.$$

According to [34], the initial values of contributions to the resonant polarization $\tilde{P}_{i1}(x,\tau=0)$ caused by different transitions, $|i\rangle \leftrightarrow |1\rangle$, $i = \{2,3,4,5\}$, are random and statistically independent. In a cylindrical active medium of the length $L$ and radius $R \ll L$, the values $\tilde{P}_{i1}(x,\tau=0)$ are defined as follows [34]. The medium is split into a set of slices $x_{k-1} \leq x < x_k$ of the length $l_{elem}$, where $x_k = k l_{elem}$, $k = 1, 2, ..., k_{max}$, and $l_{elem}$ satisfy the conditions

$$l_{elem} \leq \min_{i=2,3,4,5}\{l_{crit}^{(i)}\}, \text{ where } l_{crit}^{(i)} = \sqrt{\frac{8\pi c}{3\lambda_{21}^2 \Gamma_{rad} N_{ion} \rho_{ii}^{(0)}}}, \quad i=\{2,3,4,5\}. \tag{A3}$$

The value $l_{elem}$ should be chosen much larger than the wavelength of the resonant XUV field, $\lambda_{21} = 2\pi c/\bar{\omega}_{21}$, and much smaller than the total length of the medium, $\lambda_{21} \ll l_{elem} \ll L$. In each slice the polarization of each group of particles is determined as

$$\tilde{P}_{i1}\left((k-1)l_{elem} \leq x < k l_{elem}, \tau=0\right) = \frac{d_{tr} A_{i,k}}{\pi R^2 l_{elem}} \exp\{i\varphi_{i,k}\}, \tag{A4}$$

where the amplitudes $A_{i,k}$ and phases $\varphi_{i,k}$ are random and statistically independent. The values $A_{i,k}$ obey the following probability distribution:

$$W(A_{i,k}^2) = \frac{1}{N_{ik}} \exp(-A_{i,k}^2/N_{i,k}), \quad 0 \leq A_{i,k}^2 < \infty, \tag{A5}$$

where $N_{i,k}$ is the number of particles, which are initially excited to the state $|i\rangle$ in the slice number $k$. In the considered case of a uniform active medium $N_{i,k}$ does not depend on $k$ and can be expressed as $N_{ik} = \rho_{ii}^{(0)} N_{ion} \pi R^2 l_{elem}$. The random phases $\varphi_{i,k}$ are uniformly distributed in the interval $[0; 2\pi)$ for any $i$ and $k$:

$$W(\varphi_{i,k}) = 1/2\pi, \quad 0 \leq \varphi_{i,k} < 2\pi. \tag{A6}$$

Finally, by combing (A2) and (A4) one finds:

$$\tilde{\rho}_{i1}(x_{k-1} \leq x < x_k, \tau=0) = A_{i,k} \frac{\exp\{i(\varphi_{i,k} + \phi_i)\}}{2 N_{ion} \pi R^2 l_{elem}}, \quad i = \{2,3,4,5\}. \tag{A7}$$

The initial values of the quantum coherencies at the transitions between the excited states are set to be zero as in [22]:

$$\tilde{\rho}_{ij}(x,\tau=0)=0,\ i\neq j,\ i,j\neq 1, \tag{A8}$$

since these transitions are dipole forbidden and (in the electric dipole approximation) do not produce spontaneously emitted radiation.

## APPENDIX B. DERIVATION OF THE ANALYTICAL SOLUTION

In order to solve Eqs. (18), we use a substitution

$$\tilde{\rho}_{21}(x,\tau)=\hat{\rho}_{21}(x,\tau)e^{-\gamma_z\tau+iP_\Omega \sin\left[\Omega\tau+(1-n_{pl})\frac{\Omega}{c}x\right]}=\hat{\rho}_{21}(x,\tau)e^{-\gamma_z\tau}\sum_{k=-\infty}^{\infty}J_k(P_\Omega)e^{ik\Omega\tau}e^{ik(1-n_{pl})\frac{\Omega}{c}x}, \tag{B1}$$

where a well-known relation is implied: $\exp\{iP_\Omega \sin(\varphi)\}=\sum_{k=-\infty}^{\infty}J_k(P_\Omega)\exp(ik\varphi)$, $J_k(P_\Omega)$ is Bessel function of the first kind of order $k$, and $P_\Omega=\Delta_\Omega/\Omega$ is modulation index, which is the amplitude of variation of frequencies of the transitions $|2\rangle\leftrightarrow|1\rangle$ and $|3\rangle\leftrightarrow|1\rangle$ due to the linear Stark effect, normalized to the frequency of the modulating field. As follows from the second equation of the system (18), the function $\hat{\rho}_{21}(x,\tau)$ satisfies an equation

$$\frac{\partial \hat{\rho}_{21}}{\partial \tau}=-i\frac{d_{tr}n_{tr}}{2\hbar}E_0(x,\tau)\exp\left\{\gamma_z\tau-iP_\Omega \sin\left[\Omega\tau+(1-n_{pl})\frac{\Omega}{c}x\right]\right\}, \tag{B2}$$

which has a solution

$$\hat{\rho}_{21}=-i\frac{d_{tr}n_{tr}}{2\hbar}\sum_{m=-\infty}^{\infty}J_m(P_\Omega)\exp\left[-im\frac{\Omega}{c}(1-n_{pl})x\right]\int_0^\tau E_0(x,\tau')e^{(\gamma_z-im\Omega)\tau'}d\tau' \tag{B3}$$

valid in the case $\tilde{\rho}_{21}(x,\tau=0)=0$. One can approximately evaluate the integral in the right-hand side of Eq. (B3) as

$$\int_0^\tau E_0(x,\tau')e^{(\gamma_z-im\Omega)\tau'}d\tau'=\left\|\begin{array}{l}\Delta\tau'=\tau-\tau'\\d\Delta\tau'=d\tau'\end{array}\right\|=e^{(\gamma_z-im\Omega)\tau}\int_0^\tau E_0(x,\tau-\Delta\tau')e^{-(\gamma_z-im\Omega)\Delta\tau'}d\Delta\tau'\approx$$

$$\approx e^{(\gamma_z-im\Omega)\tau}E_0(x,\tau)\int_0^\tau e^{-(\gamma_z-im\Omega)\Delta\tau'}d\Delta\tau'=E_0(x,\tau)\frac{e^{(\gamma_z-im\Omega)\tau}-1}{\gamma_z-im\Omega}. \tag{B4}$$

Formally, this approximation implies that the amplitude of the resonant spectral component of the field, $E_0(x,\tau)$, is a slowly varying function of time at the time-interval $\sim \gamma_z^{-1}$. But as we will see from the derived solution, for the incident field with a time-independent amplitude (17) and for the values of time $\tau \geq \gamma_z^{-1}$, Eq. (B4) gives a sufficiently good approximation, since $E_0(x,\tau)$ tends to a constant value for $\tau\to\infty$. Furthermore, the initial moments of time, $\Delta\tau'\approx 0$, give the largest contribution to the integrand in (B4) due to the weigh function $e^{-\gamma_z\Delta\tau'}$, which is reduced with increasing value of $\Delta\tau'$.

Thus, one finds a solution for $\tilde{\rho}_{21}(x,\tau)$ in the form

$$\tilde{\rho}_{21}=-i\frac{d_{tr}n_{tr}}{2\hbar}E_0(x,\tau)\sum_{m,k=-\infty}^{\infty}J_m(P_\Omega)J_k(P_\Omega)\exp\left[-i(m-k)(1-n_{pl})\frac{\Omega}{c}x\right]\frac{e^{-i(m-k)\Omega\tau}-e^{ik\Omega\tau-\gamma_z\tau}}{\gamma_z-im\Omega}. \tag{B5}$$

The solution for $\tilde{\rho}_{31}(x,\tau)$ is found analogously, so that

$$\tilde{\rho}_{21} - \tilde{\rho}_{31} = -i\frac{d_{tr}n_{tr}}{2\hbar}E_0(x,\tau)\sum_{m,k=-\infty}^{\infty}\left\{1+(-1)^{m-k}\right\}J_m(P_\Omega)J_k(P_\Omega)\times$$
$$\times\frac{e^{-i(m-k)\Omega\tau}-e^{ik\Omega\tau-\gamma_z\tau}}{\gamma_z-im\Omega}\exp\left\{-i(m-k)(1-n_{pl})\frac{\Omega}{c}x\right\}. \quad (B6)$$

In the following, we assume that the frequency of the modulating field is much larger than the linewidth of the resonant transition, $\Omega/\gamma_z \gg 1$, so that for $J_0(P_\Omega) \neq 0$ a contribution with $m=0$ is dominant in Eq. (B6), which is reduced to

$$\tilde{\rho}_{21} - \tilde{\rho}_{31} = -i\frac{d_{tr}n_{tr}}{\hbar\gamma_z}E_0(x,\tau)(1-e^{-\gamma_z\tau})J_0(P_\Omega)\sum_{l=-\infty}^{\infty}J_{2l}(P_\Omega)e^{-i2l\Omega\tau}\exp\left\{-i2l(1-n_{pl})\frac{\Omega}{c}x\right\}. \quad (B7)$$

Thus, according to (18), z-polarized XUV field is determined by the equation

$$\frac{\partial \tilde{E}_z}{\partial x} = \frac{4\pi\omega_z N_{ion}n_{tr}d_{tr}^2}{\hbar c\sqrt{\varepsilon_{X-ray}}\gamma_z}E_0(x,\tau)(1-e^{-\gamma_z\tau})J_0(P_\Omega)\sum_{l=-\infty}^{\infty}J_{2l}(P_\Omega)e^{-i2l\Omega\tau}e^{-i2l(1-n_{pl})\frac{\Omega}{c}x}, \quad (B8)$$

which has a solution in the form of the spectral comb (15), that is, $\tilde{E}_z(x,\tau) = \sum_{l=-\infty}^{\infty}E_l(x,\tau)e^{-i2l\Omega\tau}$.

The amplitudes of the spectral components of z-polarized field (15), which satisfy the boundary condition (17), are given by equations (19) of the paper.

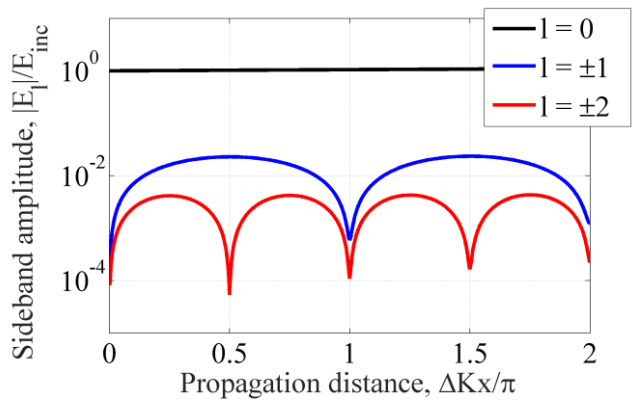

Fig. 1. (Color online) Amplitudes of 0th, ±1st and ±2nd spectral components of the XUV field vs. the propagation distance (on a horizontal axis we plot the phase incursion of the modulating field due to plasma dispersion, $\Delta Kx$, normalized by $\pi$) plotted according to (19) for $g_0/\Delta K = 13.8$ and $P_\Omega = 3.3$, and for the very initial moment of time, $\gamma_z \tau = 0.01$. The amplitudes of sidebands oscillate and reach their maxima at (i) $\Delta Kx = \pi/2 + n\pi$, where $n=1,2,...$, for the ±1 spectral components and (ii) at $\Delta Kx = \pi/4 + n\pi/2$ for the ±2 spectral components in agreement with (22), while the resonant (0th) spectral component is almost not amplified. The sidebands are at least two orders of magnitude weaker than the resonant spectral component of the field.

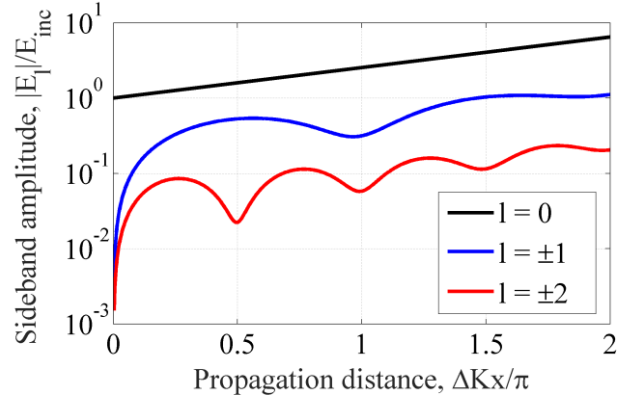

Fig. 2. (Color online) Same as in Fig. 1, but for $\gamma_z \tau = 0.2$. With increasing time left from the beginning of interaction the resonant response is intensified, so that the amplitudes of sidebands grow up to approximately 1/10 of the resonant spectral component.

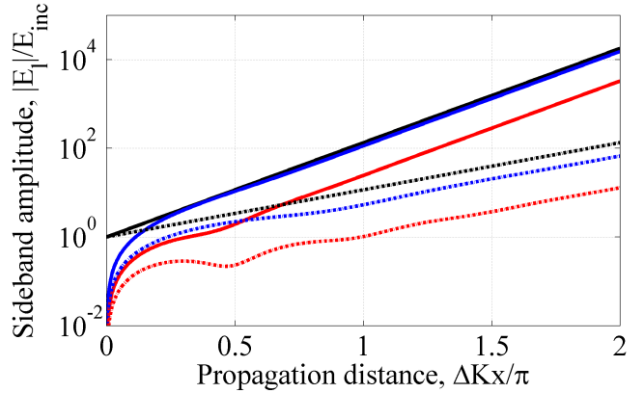

Fig. 3. (Color online) Same as in Figs. 1 and 2, but for $\gamma_z \tau = 3$. Solid lines correspond to $g_0/\Delta K = 13.8$ (as in Figs. 1 and 2); dotted lines are plotted for $g_0/\Delta K = 6.9$. In the steady-state regime, $\tau \gg \gamma_z^{-1}$, the amplitudes of sidebands reach their maximum values and might be comparable with the amplitude of the resonant spectral component.

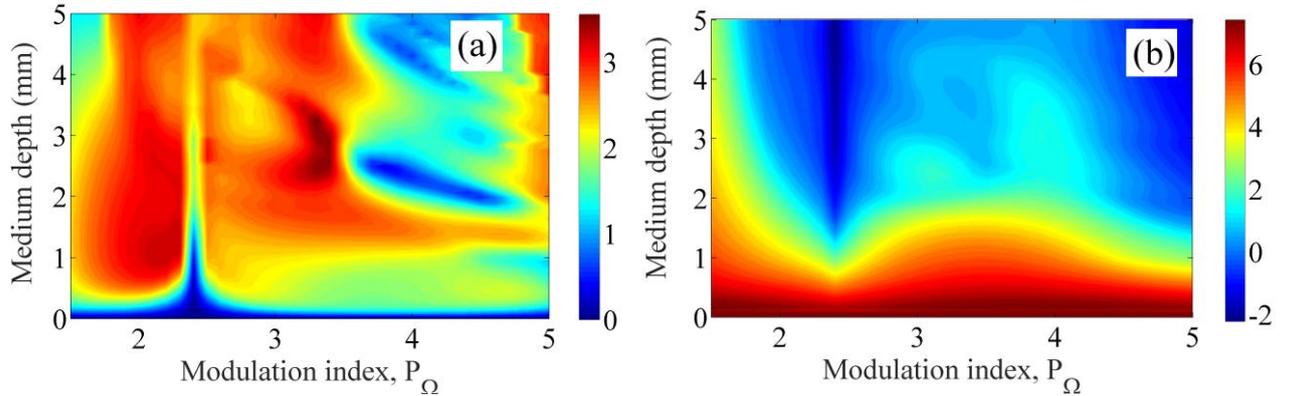

Fig. 4. (Color online) (a) The dependence of contrast of the produced pulses on the propagation distance through the medium, $x$, and on the modulation index, $P_\Omega$. (b) The ratio between peak intensities of z-polarized and y-polarized components of the output XUV field shown on a logarithmic scale as a function of $x$ and $P_\Omega$.

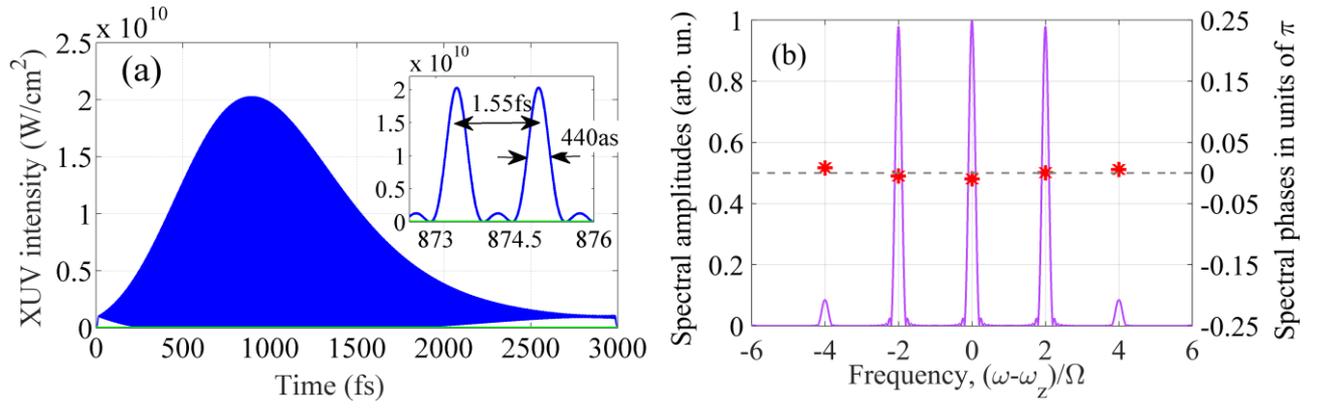

Fig. 5. (Color online) (a) The time dependence of intensity of the XUV field for $x=1.1$ mm and $P_\Omega=2.2$. The parameters of the modulating IR field are $\lambda_{IR}=930$ nm and $I_{IR}=4\times10^{14}$ W/cm$^2$. The seeding XUV field has the intensity $I_{inc}=10^9$ W/cm$^2$ and duration 3 ps. Blue line shows intensity of z-polarized XUV field, green line corresponds to y-polarized field. The ratio between peak intensities of z- and y-polarization components of the XUV field is 1600/1. (b) The windowed spectrum of z-polarized XUV field near the peak of the pulse train in panel (a). Lavender curve represents the amplitudes of spectral components, while red stars show the spectral phases at their carrier frequencies (the linear dependence of the phase on the frequency has been removed).

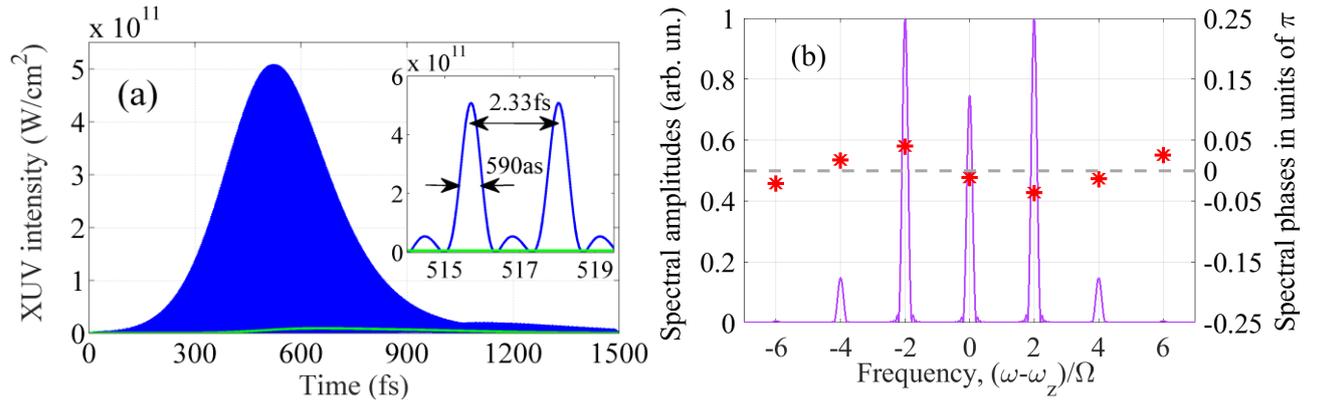

Fig. 6. (Color online) Same as in Fig. 5, but for $x=2.5$ mm and $P_\Omega=3.3$. The modulating field has the wavelength $\lambda_{IR}=1400$ nm and intensity $I_{IR}=4\times10^{14}$ W/cm$^2$. The seeding XUV field has the intensity $I_{inc}=10^9$ W/cm$^2$ and duration 1.5 ps. The ratio between peak intensities of z- and y-polarization components of the output XUV field is 30/1.

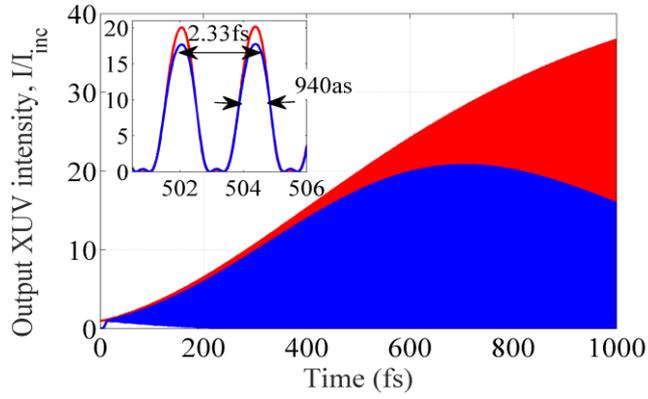

Fig. 7. (Color online) A comparison of the analytical (red) and numerical (blue) solutions for the intensity of the amplified z-polarized XUV field. The gain coefficient is two times smaller than that assumed in Fig. 1-6, except dotted lines in Fig. 3. The parameters of the medium and the modulating field are $x$=1.15 mm, $\lambda_{IR}$=1400 nm, $I_{IR}$=4×10$^{14}$ W/cm$^2$, $P_\Omega$=3.3, $g_0/\Delta K$ = 6.9. The output XUV intensity is normalized to the intensity of the seeding field, $I_{inc}$=10$^9$ W/cm$^2$.